\documentclass[journals,article,submit,pdftex,moreauthors]{Definitions/mdpi}

\firstpage{1} 
\pubvolume{1}
\issuenum{1}
\articlenumber{0}
\pubyear{2024}
\copyrightyear{2024}
\datereceived{ } 
\daterevised{ } 
\dateaccepted{ } 
\datepublished{ } 

\usepackage{inputenc}
\usepackage{listings}
\usepackage{xcolor}
\usepackage{framed}
\usepackage{adjustbox}
\usepackage{longtable}
\usepackage{geometry}
\usepackage{tabularx}
\usepackage{mdframed}
\usepackage{array}
\usepackage{cellspace}
\usepackage{setspace}
\definecolor{shadecolor}{rgb}{0.95, 0.95, 0.95}
\usepackage{graphicx}  
\usepackage{caption}   

\hreflink{https://doi.org/} 

\Title{Detecting Zero-Day Web Attacks with an Ensemble of LSTM, GRU, and Stacked Autoencoders}

\TitleCitation{Title}

\Author{Vahid Babaey $^{1}$\orcidA{} and Hamid Reza Faragardi$^{2}$ }

\AuthorNames{Vahid Babaey Lastname and Hamid Reza Faragardi}

\AuthorCitation{Babaey, V.; Faragardi, H.}

\address{%
$^{1}$ \quad Department of Electrical and Computer Engineering, University of North Carolina at Charlotte; vbabaey@charlotte.edu\\
$^{2}$ \quad Research Engineer, KTH Royal Institute of Technology, Stockholm, Sweden; Hamid.faragardi@trendplus.se}

\corres{Correspondence: vbabaey@charlotte.edu, Hamid.faragardi@trendplus.se}

\modulolinenumbers[0] 
\nolinenumbers
\abstract{
The rapid growth in web-based services has significantly increased security risks related to user information, as web-based attacks become increasingly sophisticated and prevalent. Traditional security methods frequently struggle to detect previously unknown (zero-day) web attacks, putting sensitive user data at significant risk. Additionally, reducing human intervention in web security tasks can minimize errors and enhance reliability.
This paper introduces an intelligent system designed to detect zero-day web attacks using a novel one-class ensemble method consisting of three distinct autoencoder architectures: LSTM autoencoder, GRU autoencoder, and stacked autoencoder. Our approach employs a novel tokenization strategy to convert normal web requests into structured numeric sequences, enabling the ensemble model to effectively identify anomalous activities by uniquely concatenating and compressing the latent representations from each autoencoder.
The proposed method efficiently detects unknown web attacks while effectively addressing common limitations of previous methods, such as high memory consumption and excessive false positive rates. Extensive experimental evaluations demonstrate the superiority of our proposed ensemble, achieving remarkable detection metrics: 97.58\% accuracy, 97.52\% recall, 99.76\% specificity, and 99.99\% precision, with an exceptionally low false positive rate of 0.2\%. These results underscore our method’s significant potential in enhancing real-world web security through accurate and reliable detection of web-based attacks.
}


\usepackage{tabularx}
\usepackage{ragged2e}
\makeatletter
\def\@mylogo{}
\makeatother
\begin{document}
\modulolinenumbers[0] 
\nolinenumbers

\section{Introduction}
\label{sec:introduction}

In modern digital infrastructure, websites and web-based applications play a crucial role in facilitating economic, educational, recreational, and political activities. However, as the reliance on these platforms increases, so does the risk of security threats, including unauthorized access, data breaches, and service disruptions. One of the primary attack vectors involves manipulating web requests, where adversaries masquerade as legitimate users to exploit vulnerabilities. Consequently, the detection and mitigation of malicious web requests have become vital for ensuring the security of any online service, including websites, web applications, and Content Delivery Networks (CDNs).

To counter such threats, various security mechanisms, including Web Application Firewalls (WAFs) and blacklisting techniques, have been deployed. While these methods offer some level of protection, they remain ineffective against zero-day attacks—novel exploits that lack predefined security signatures~\cite{ahmad2023zero}. The primary challenge associated with zero-day attacks lies in their unpredictability, as they introduce previously unseen patterns that traditional rule-based detection systems fail to recognize. Addressing these challenges through deep learning-based anomaly detection presents a promising approach, leveraging neural networks to autonomously identify deviations indicative of malicious activity.

Conventional methods for preventing web-based attacks, such as WAFs~\cite{dawadi2023deep} and blacklisting, exhibit several limitations. For instance, maintaining a blacklist of prohibited keywords within web requests is both time-consuming and insufficient in addressing evolving attack patterns. Moreover, none of these existing approaches is capable of detecting zero-day attacks, as the strategies and obfuscation techniques employed in these attacks remain unknown. Recent advancements in machine learning and deep learning have demonstrated significant potential in enhancing security through intelligent threat identification, making these techniques highly relevant for modern cyber defense systems.

A critical advantage of anomaly detection models is that they do not require prior exposure to zero-day attacks to effectively detect them. In this study, an ensemble model is proposed that integrates multiple sub-models designed to detect zero-day attacks. Given that the patterns of zero-day attacks are inherently unknown, the model is trained exclusively on normal web request data. By learning the distribution of normal web traffic, the model becomes proficient in identifying deviations, thereby flagging both known and previously unseen attacks as anomalous. This approach ensures that malicious requests, whether originating from known attack types or zero-day exploits, are effectively classified as security threats.

To evaluate the proposed system, various web attacks such as SQL Injection (SQLi), Cross-Site Scripting (XSS), and Buffer Overflow~\cite{yang2023systematic} are treated as zero-day attacks within the dataset. The model classifies any request with an anomaly score exceeding a predefined threshold as a potential zero-day attack. While the proposed approach does not explicitly categorize different types of attacks, it demonstrates the capability to reliably detect anomalous activities, ensuring a high level of security against emerging threats. The primary objective of this model is to simultaneously address both known and zero-day attacks while maintaining a high detection rate and minimizing false positives.

The rest of this paper is structured as follows: Section~\ref{sec:background} presents the foundational concepts and research background. Section~\ref{sec:related_work} provides a review of existing literature on web attack detection. The methodology and architectural design of the proposed model are discussed in Section~\ref{sec:Proposed model}, followed by a performance evaluation in Section~\ref{sec:evaluation_results}. Section~\ref{sec:discussion} elaborates on the broader implications of the findings, and Section~\ref{sec:conclusions} concludes the paper with final remarks.

The key contributions of this research are as follows: 

\begin{itemize}
    \item \textbf{Innovative Ensemble Model Architecture:} This study introduces a novel ensemble approach by integrating LSTM, GRU, and stacked autoencoders for anomaly detection in web requests. Unlike conventional ensemble methods that use simple averaging or majority voting, our approach uniquely concatenates and compresses the latent representations from each autoencoder. This technique significantly improves anomaly detection performance and computational efficiency.
    \item \textbf{Advanced Tokenization and Feature Mapping:} We propose a novel tokenization strategy that classifies tokens based on their character composition (numeric, lowercase, uppercase, and special characters). This structured approach effectively reduces input dimensionality, ensures greater consistency in data representation, and significantly enhances the detection capability of our anomaly detection system.
    \item \textbf{Zero-Day Attack Detection:} Our model is trained exclusively on normal web requests, enabling it to effectively identify and detect previously unseen zero-day attacks by capturing deviations from established normal request patterns.
    \item \textbf{Comprehensive Evaluation Metrics with Emphasis on False Positive Rate (FPR):} Unlike many existing studies, we explicitly evaluate and report the False Positive Rate, achieving a significantly lower FPR of 0.2\%. This comprehensive evaluation underscores the practical applicability of our model, addressing an essential aspect often overlooked in anomaly detection research.
\end{itemize}

By addressing the limitations of traditional detection systems and leveraging anomaly detection through deep learning, this research contributes to advancing cybersecurity measures against evolving web-based threats.

\section{Background}
\label{sec:background}

Due to the increasing reliance on internet-based services, web attacks pose significant threats to user privacy and can severely disrupt web server operations, affecting users on a global scale. Among these threats, zero-day (previously unknown) attacks are particularly concerning, as they can lead to privacy breaches and denial of service, impacting all users of a targeted website~\cite{calzavara2020machine}. There are two approaches to detect attacks: non-ML (heuristic-based) and machine learning-based approaches. Heuristic methods play a crucial role in cybersecurity due to their ability to rapidly identify potentially malicious activities based on predefined rules and patterns, without requiring extensive labeled datasets. Heuristic algorithms analyze data patterns and system behaviors, identifying threats by matching observed behaviors against known suspicious patterns or predefined rules. The primary advantage of heuristic approaches lies in their capability to detect novel or zero-day threats promptly, often faster than traditional signature-based methods. However, heuristic approaches can suffer from high false positives or negatives due to their reliance on manually defined rules and parameters, and thus require continuous refinement and adaptation. Despite these limitations, heuristic algorithms remain widely adopted in cybersecurity for their interpretability, speed, and ability to quickly detect anomalies in real-time environments \cite{10912106}. Machine learning-based approaches for web attack detection typically consist of two key phases: training and detection. During the training phase, the model learns from patterns of normal web requests, while in the detection phase, it utilizes this learned knowledge to identify and mitigate potential web attacks~\cite{ahmed2016survey}. 

Web attack countermeasures can generally be categorized into three primary approaches: (1) supervised, (2) unsupervised, and (3) semi-supervised learning. The supervised approach is primarily designed to detect known attacks and is commonly implemented in signature-based systems such as Web Application Firewalls (WAFs). These models rely on labeled datasets containing both historical attack patterns and normal requests, making them highly effective against previously documented threats. However, their reliance on predefined signatures renders them ineffective against zero-day attacks, as these attacks introduce novel patterns that are not represented in the training dataset~\cite{pu2020hybrid}.

The unsupervised approach, on the other hand, employs anomaly detection techniques to distinguish between normal system activities and anomalies. Unlike supervised methods, this approach does not rely on historical attack patterns, allowing it to identify previously unseen zero-day attacks~\cite{pu2020hybrid,long2020efficient}. By modeling the expected behavior of normal web traffic, anomaly detection-based methods can effectively flag deviations indicative of malicious activity, making them particularly suitable for dynamic and evolving attack landscapes.

The semi-supervised approach, which lies between supervised and unsupervised methods, leverages normal web request data to train the model. This approach focuses exclusively on learning the characteristics of legitimate requests, enabling the model to differentiate between normal and anomalous activities. Since it involves training on only one category of data—normal requests—it eliminates the need for labeled attack samples while still allowing for anomaly detection in the detection phase~\cite{pu2020hybrid}.

In this research, an unsupervised approach is employed to address the challenge of detecting zero-day attacks, which lack predefined signatures and attack patterns. By learning the distribution of normal web requests, the proposed model can effectively identify requests that deviate from the established normal behavior, thereby flagging them as potential zero-day attacks. This approach enhances the system's ability to detect novel and previously unknown threats, making it a robust solution for mitigating modern web security risks.

\section{Related Work}
\label{sec:related_work}

Numerous methods and models have been proposed to counter web attacks, including zero-day attacks. Models by researchers like Pu et al. \cite{pu2020hybrid}, Ingham et al. \cite{ingham2007learning}, Sivri et al. \cite{sivri2022web}, Jung et al. \cite{jung2021pf}, Vartouni et al. \cite{vartouni2018anomaly}, Ariu et al. \cite{ariu2011hmmpayl}, Liang et al. \cite{liang2017anomaly}, Kuang et al. \cite{kuang2019deepwaf}, Tang et al. \cite{tang2020zerowall}, Indrasiri et al. \cite{indrasiri2021robust}, Gong et al. \cite{gong2021model}, Tekerek et al. \cite{tekerek2021novel}, Jemal et al. \cite{jemal2022swaf}, Alaoui et al. \cite{alaoui2022web}, Mohamed et al. \cite{moarref2023mc}, Yuan et al \cite{app132111763}, Vorobyov et al. \cite{10.1145/3639476.3639772}, Su et al. \cite{10.1145/3597926.3598116}, Silvestre et al. \cite{enase24}, Yatagha et al. \cite{yatagha2024towards}, Katbi et al. \cite{katbi2025one}, Tokmak et al. \cite{tokmak2023stacking}, Alqhwazi et al. \cite{alghawazi2023deep}, Thalji et al. \cite{thalji2023ae}, Yao et al. \cite{yao2023lightweight}, are notable.

The model by Pu et al. \cite{pu2020hybrid} introduces an unsupervised anomaly detection method that combines Sub-Space Clustering (SSC) and One Class Support Vector Machine (OCSVM). This approach aims at detecting cyber intrusions without prior knowledge of the attacks, making it particularly useful for identifying unknown or zero-day attacks.

The model by Ingham et al. \cite{ingham2007learning} proposes a method for detecting web attacks by focusing on deep learning techniques, specifically utilizing Transformer models. This approach represents a major advancement in web security, providing a more dynamic and intelligent system for detecting and mitigating web-based attacks.

Sivri et al. \cite{sivri2022web} used various machine learning and deep learning models. For example, a hybrid model that uses character-level representations to classify HTTP requests as normal or malicious. The study employed models such as XGBoost, LightGBM, LSTM and CNN. The upsampling techniques have been used to balance the dataset, which helps improve classification metrics. As a result, the LSTM model achieved the best accuracy and F1 score, while LightGBM performed better in computation time. This work shows the importance of balancing in real-time web intrusion detection systems.

Jung et al. \cite{jung2021pf} used a novel approach named Payload Feature-Based Transfer Learning (PF-TL) to cope with insufficient training data in intrusion detection systems. Their method leverages knowledge transfer from a labeled source domain to an unlabeled target domain by extracting features from both the header and payload of network traffic. The technique they use is a hybrid feature extraction, combining signature-based and text vectorization methods, to enhance the representation of attack patterns.

The model by Vartouni et al. \cite{vartouni2018anomaly} uses a deep neural network-based method for feature learning and isolation forest for classification to identify malicious requests. It employs an n-gram model, which represents overlapping subsets of n characters from the data. 

Ariu et al. \cite{ariu2011hmmpayl} model is an intrusion detection system that represents payloads as byte sequences, analyzed using Hidden Markov Models (HMM). This proposed algorithm ensures the analytical power of n-gram analysis while overcoming its computational complexity, using HMM for feature extraction. However, HMM models are less effective when the sequence length is not appropriate, leading to poorer performance in processing complex requests.

In the Liang et al. \cite{liang2017anomaly} model, the approach involves first training two Recurrent Neural Networks (RNNs) with Complex Recurrent Units (LSTM or GRU units) to learn normal request patterns solely from unsupervised normal requests. Then, a supervised neural network classifier is trained, taking the output of the RNN as input to categorize normal and abnormal requests.

Kuang et al. \cite{kuang2019deepwaf} employ deep learning concepts to design a model named DeepWaf, a combination of LSTM and CNN deep neural networks, achieving satisfactory results in detecting web attacks.

In the Tang et al. \cite{tang2020zerowall} model, each word in an HTTP request (except for low-value words like 'and', 'or', etc.) is tokenized. Words and tokens are mapped to each other through TokenIDs. The tokenized request is then encoded and decoded using a Short-Term Memory architecture; if the decoded value matches the pre-encoded tokenized value, the request is benign, otherwise, it's malicious. The model primarily targets zero-day attacks, leaving known attack detection to WAF and addressing zero-day attacks through the Zero-Wall model. A limitation of this approach is that new benign requests with different patterns might be incorrectly flagged as malicious and subjected to further scrutiny in the Zero-Wall model before being identified as benign.

In the Indrasiri et al. \cite{indrasiri2021robust} model, seven classification algorithms, one clustering algorithm, two ensemble methods, and two large standard datasets with 73,575 and 100,000 URLs were used. Two testing modes (percentage split, K-Fold cross-validation) were employed for experiments and predictions. An ensemble model named ERG-SVC was proposed, using features selected by various feature selection methods.

The model by Gong et al. \cite{gong2021model} proposes a method to improve web attack detection by incorporating model uncertainty into deep learning (DL) models, specifically focusing on Convolutional Neural Networks (CNNs). The method aims to address the problem of annotation errors in training data. Annotation errors are common in web attack datasets due to the vast and varied nature of web traffic, making correct labeling challenging.

The model by Tekerek et al. \cite{tekerek2021novel} introduces a novel approach for detecting web-based attacks using a deep learning architecture centered on Convolutional Neural Networks (CNNs). Focused on anomaly-based detection, this method preprocesses HTTP request data, particularly URLs and payloads, to identify unusual patterns indicative of potential threats.

The model by Jemal et al. \cite{jemal2022swaf} presents a smart web application firewall (SWAF) based on a convolutional neural network. The model is evaluated using a 5-fold cross-validation method. The CNN is characterized by a specific architecture. It can process data at scale and automatically extract and select features and consists of five layers.

The model by Alaoui et al. \cite{alaoui2022web} proposes an approach based on Word2vec embedding and a stacked generalization ensemble model for LSTMs to detect malicious HTTP web requests.

Mohamed et al. \cite{mohamed2023multi} propose a deep learning-based multi-class intrusion detection system that classifies different types of web attacks using algorithms like LSTM, Bi-LSTM, CNN, and RNN. Automatic extraction and classification of features from HTTP traffic are their main approaches which overcome limitations related to traditional feature engineering.

The model by Shahid et al. \cite{shahid2022enhanced} proposes a framework based on an enhanced hybrid approach where Deep Learning model is nested with a Cookie Analysis Engine for web attacks detection, mitigation and attacker profiling in real time.

The model by Moarref et al. \cite{moarref2023mc} tries to focus on enhancing web attack detection via a character-level multichannel multilayer dilated convolutional neural network, processes HTTP request texts at the character level and extract relevant features. The model combines multichannel dilated convolutional blocks with varying kernel sizes to capture diverse temporal relationships and dependencies among characters.

Yatagha et al. \cite{yatagha2024towards} proposed a hybrid anomaly detection model combining VAE, LSTM, and OCSVM to detect zero-day anomalies in cyber-physical systems. The model learns normal patterns and flags deviations using reconstruction errors and latent space analysis. An adaptive loss adjustment algorithm ensures continual learning without forgetting. Deployed on a Raspberry Pi, the system effectively detects contextual anomalies in real-time.

Katbi et al. \cite{katbi2025one} proposed IDSVDD, a novel one-class anomaly detection framework for IoT environments that combines Deep SVDD with an interpolated adversarial autoencoder. The model enhances the structure of the latent space by enforcing convexity and regularization through adversarial interpolation, making it easier to distinguish anomalies from normal data. By learning a compact hypersphere that encloses only normal samples, the system achieves strong zero-day detection performance across multiple IoT datasets while remaining lightweight enough for deployment in resource-constrained environments.

Tokmak et al. \cite{tokmak2023stacking} presented a deep learning framework for zero-day threat detection that combines Stacked Autoencoders (SAE) for feature selection with an LSTM classifier. Using the UGRansome dataset, the model first performed unsupervised feature extraction with SAE, then applied supervised LSTM layers to capture temporal patterns. The hybrid SAE-LSTM model achieved high accuracy (98\%) across signature, anomaly, and synthetic signature attacks, showing strong generalization and effectiveness for detecting both known and novel threats.

Alqhwazi et al. \cite{alghawazi2023deep} proposed an SQL injection detection system using a Recurrent Neural Network (RNN) Autoencoder, trained on a public Kaggle dataset of SQL queries. Their architecture uses an autoencoder for dimensionality reduction, followed by an LSTM layer for classification. The model achieved 94\% accuracy and 92\% F1-score, outperforming traditional ML classifiers like SVM, decision tree, and logistic regression. This approach effectively captures long-term dependencies in SQL queries, making it well-suited for detecting complex or obfuscated injection attacks.

Thalji et al. \cite{thalji2023ae} proposed AE-Net, a novel autoencoder-based feature engineering approach for detecting SQL injection attacks. AE-Net extracts high-level deep features from SQL textual queries, which are then used as input to multiple machine learning and deep learning models. Among the models tested, Extreme Gradient Boosting (XGBoost) achieved the highest performance, reaching a 0.99 accuracy score in k-fold cross-validation. The approach demonstrated the effectiveness of deep unsupervised feature learning in enhancing SQLi detection over traditional methods like BoW and TF-IDF.

Yao et al. \cite{yao2023lightweight} proposed a lightweight intrusion detection system for IoT that combines a One-Class Bidirectional GRU Autoencoder with Soft-Voting Ensemble Learning. The autoencoder is trained on only normal data to detect anomalies—including zero-day attacks—based on reconstruction loss. Detected anomalies are then classified using an ensemble of Random Forest, XGBoost, and LightGBM to identify the closest known attack type. The system demonstrated high accuracy and adaptability across three benchmark datasets: WSN-DS, UNSW-NB15, and KDD99.

Apart from machine learning techniques, various heuristic-based approaches have been proposed to detect web-based attacks, such as Yuan et al. \cite{app132111763} who proposed a static SQL injection detection technique based on program transformation to address the limitations of existing tools in handling object-oriented database extensions (OODBE) in PHP applications. Their method, OODBE-SCAN, first transforms object-oriented constructs into semantically equivalent procedural code, enabling accurate identification of source and sink points. The method then performs control flow graph construction and taint analysis to detect vulnerabilities. Compared to tools like RIPS and Seay, OODBE-SCAN demonstrated superior precision and recall in detecting real-world vulnerabilities in OODBE-based web applications.

Vorobyov et al. \cite{10.1145/3639476.3639772} introduced a novel runtime protection mechanism against SQL injection attacks based on synthesizing fine-grained allowlists from benign SQL queries. Their approach uses an information flow model to decompose SQL queries into semantic units called information tuples, which capture disclosed columns, accessed fields, and related predicates. By generalizing these tuples across a set of safe queries, they create a context-sensitive allowlist that permits future queries only if they disclose no more information than allowed. This method outperforms syntax-based detectors by focusing on semantic disclosure rather than structural similarity, thus reducing false positives and better preventing data exfiltration.

Su et al. \cite{10.1145/3597926.3598116} proposed Splendor, a static analysis framework for detecting stored Cross-Site Scripting (XSS) vulnerabilities in modern PHP web applications, especially those using Data Access Layers (DAL). The approach introduces a fuzzy token-matching technique to identify database operation triples (table, column, and operation type) from code fragments, even when SQL queries are dynamically constructed or obscured through encapsulation. Splendor then performs a two-phase taint analysis, tracing tainted data from sources to the database writes and from the database reads to sinks. The framework demonstrated strong scalability, identifying 17 real-world zero-day vulnerabilities and outperforming both static (RIPS) and dynamic (Black Widow) tools.

Silvestre et al. \cite{enase24} introduced FreeSQLi, a novel static analysis tool that detects SQL injection vulnerabilities in PHP applications using session types. Their approach translates PHP code into the FreeST programming language, which supports rich type systems modeling communication protocols. By interpreting interactions between the application and the database as typed communication sessions, FreeSQLi checks for type mismatches—such as sending unsanitized user input (typed as Unsafe) to sensitive sinks—and flags these inconsistencies as SQLi vulnerabilities. This method offers formal guarantees and reduces false positives by leveraging strong type checking rather than heuristics or machine learning.

A key limitation of the related works is the omission of one of the most critical evaluation metrics: the False Positive Rate (FPR), which quantifies the proportion of normal web requests misclassified as malicious. This metric is essential for assessing the model’s capability to accurately encode and decode normal requests, ensuring minimal disruption to legitimate traffic. In contrast, our proposed model demonstrates superior performance, achieving the lowest False Positive Rate of 0.2\%, which is even lower than the values reported in prior studies that included this metric in their evaluation.

\section{Proposed model}
\label{sec:Proposed model}

This section presents the proposed model for detecting zero-day web attacks. The detection process begins with the pre-processing of web requests through tokenization techniques, ensuring standardized input representation. These processed requests are then fed into an ensemble of relatively simple one-class classifiers designed to distinguish between normal and malicious web traffic. The effectiveness of the proposed model is assessed during the detection phase, focusing on its capability to identify and mitigate advanced security threats.

\subsection{Architecture}

\begin{figure}[htbp]
\centerline{\includegraphics[width=1\linewidth, height=9.5cm]{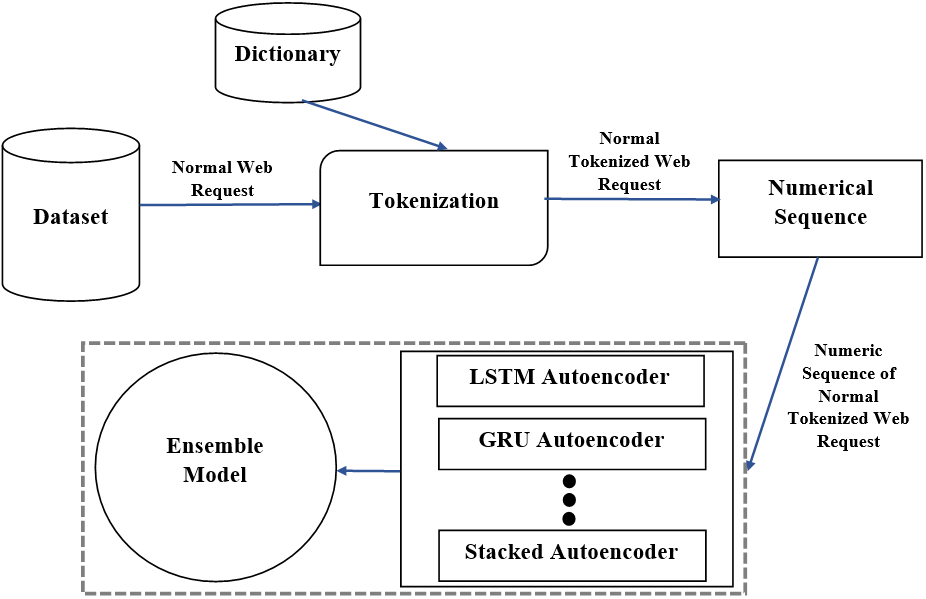}} 
\caption{The proposed model in the training phase}
\label{fig1} 
\end{figure}

\begin{figure}[htbp]
\centerline{\includegraphics[width=1\linewidth, height=9.5cm]{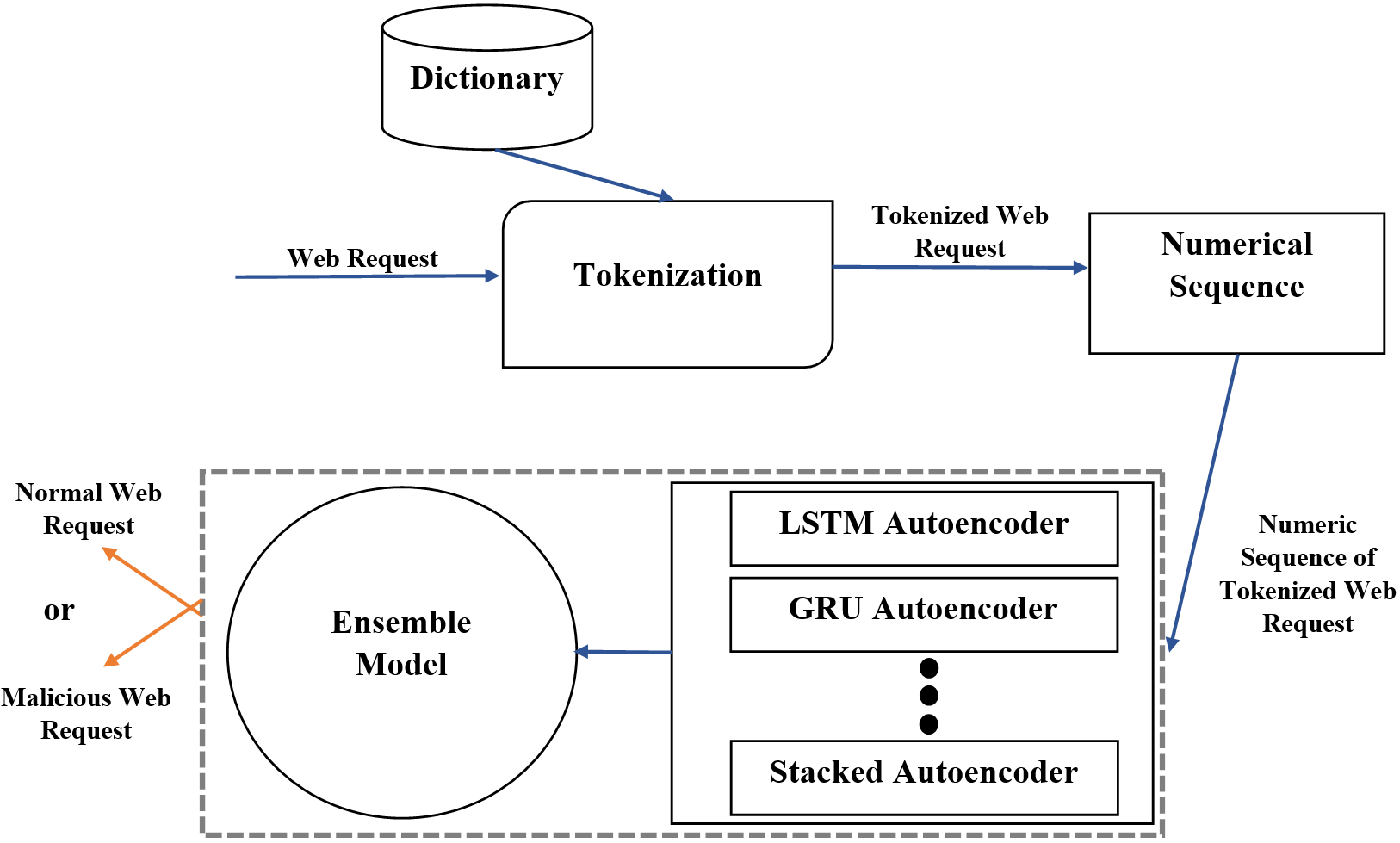}} 
\caption{The proposed model in the test phase}
\label{fig2} 
\end{figure}

The proposed model comprises multiple components, each serving a distinct role in the detection process. These components work together to predict whether an incoming web request is benign or malicious. The overall model architecture is depicted in Figure~\ref{fig1} for the training phase and Figure~\ref{fig2} for the testing phase.

During the training phase, the model is exclusively trained on normal web requests to establish a baseline pattern of legitimate traffic. In this phase, the model has full access to the training dataset, allowing it to learn the distribution of normal request patterns~\cite{thomas2022dynamic}. Conversely, in the testing phase, both normal and malicious web requests are input into the model for evaluation. This enables the model to assess its ability to generalize and detect deviations indicative of potential zero-day attacks.

\subsubsection{Tokenization}

A key innovation of the proposed model is the introduction of a novel tokenization technique for web requests, applied at the word level to both normal and malicious inputs~\cite{khalid2019predicting}. Due to the inherent variability in the length and structure of web requests, this method addresses the challenges of training neural network-based models for web security, which arise from the inherent variability in the length and structure of web requests. Specifically, the model leverages anomaly detection principles to effectively distinguish between legitimate and anomalous web traffic~\cite{levene2004learning}. 

To ensure consistent data representation, the pre-processing pipeline standardizes normal web requests through a dictionary-based tokenization approach~\cite{liang2017anomaly}. This process involves segmenting each request at the word level using tools such as Python's WordPunctTokenizer~\cite{vijayarani2016text}. The resulting structured pattern is subsequently utilized as the input for training the ensemble model. We will explain the tokenization and data processing workflow by applying it to an example from our dataset. The example request is as follows:\\
\texttt{POST /tienda1/publico/registro.jsp?modo=registro\&login=m6\&password=m6\\
\&nombre=m\&apellidos=m\&email=m\&dni=mm\&direccion=Calle+Salvatierra+196+\\
\%2C+\&ciudad=m\&provincia=31\&cp=68970\&ntc=6987987070987097\&B1=Registrar}



\begin{figure}[htbp]
\centerline{\includegraphics[width=1\linewidth, height=4.5cm]{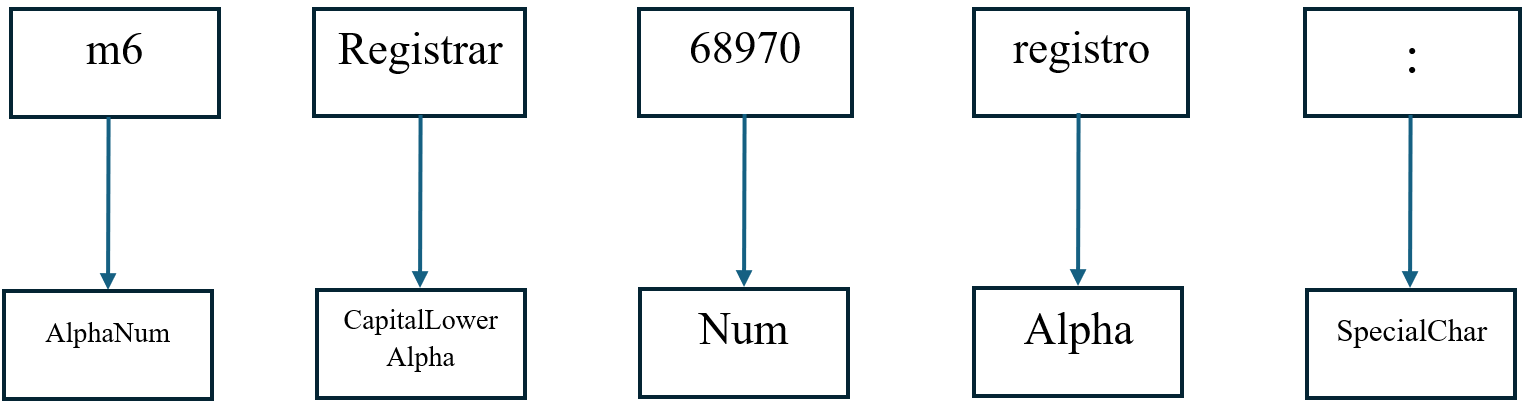}} 
\caption{A Tokenized request.}
\label{fig3} 
\end{figure}

A predefined dictionary is utilized to categorize each character into distinct classes, facilitating structured tokenization. The dictionary includes categories such as \textit{Alpha}, \textit{AlphaNum}, \textit{CapitalAlpha}, and \textit{SpecialChar}, among others. For instance, in Figure~\ref{fig3}, the token "login" is assigned the value "m6," which represents a combination of letters and numbers, classifying it under \textit{AlphaNum} according to the predefined dictionary. Similarly, the word "Register," which begins with a capital letter, falls under the \textit{CapitalLowerAlpha} category. These classifications enhance the accuracy of tokenization and facilitate the interpretation of web requests within the proposed model.

The significance of the tokenization approach in this model is twofold:

\begin{itemize}
    \item \textbf{Data volume reduction:} The tokenization process optimizes data representation, reducing the complexity and size of input data, thereby improving computational efficiency.
    \item \textbf{Pattern identification for anomaly detection:} By establishing a structured pattern for normal web requests, the tokenization method enhances the model’s ability to differentiate between legitimate and anomalous activities, ensuring higher accuracy in detecting malicious web requests.
\end{itemize}

\subsubsection{Numerical Sequence}

\begin{figure}[htbp]
\centerline{\includegraphics[width=1\linewidth, height=7cm]{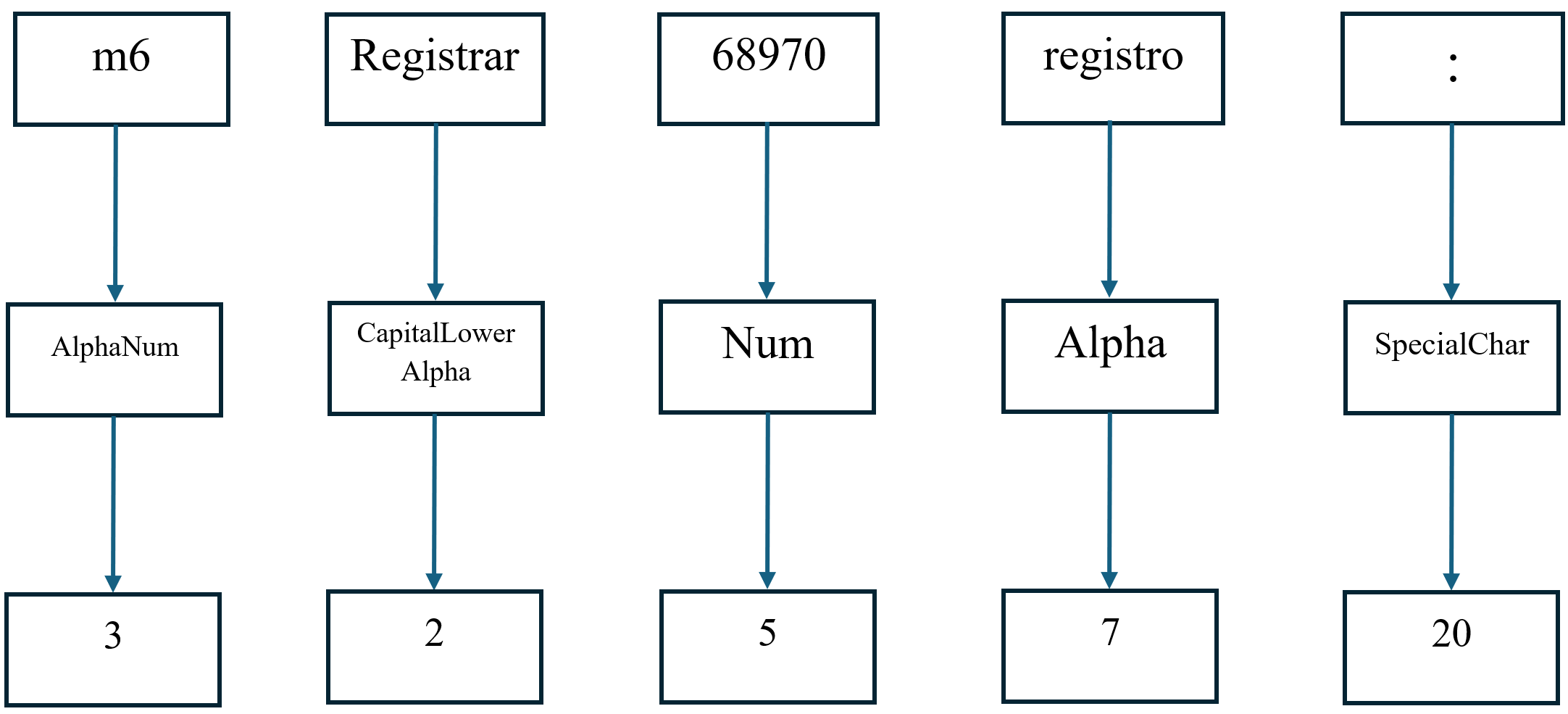}} 
\caption{Mapping words to numbers.}
\label{fig4} 
\end{figure}

Following the tokenization of web requests on a word-by-word basis, each token must be mapped to a corresponding numerical value, as neural networks operate on numbers rather than raw text. This transformation is a critical step, as the varying range of input features necessitates data scaling before being processed by the model~\cite{rashvand2023real}. The tokenized text is converted into a structured numerical format suitable for input into the neural network. Figure~\ref{fig4} illustrates the mapping process, where individual words are assigned numerical values, facilitating seamless integration into the model.

To accommodate variations in the length of numerical sequences representing web requests, a padding mechanism is implemented. This ensures that all input sequences maintain a uniform length, preventing inconsistencies in model processing. Padding standardizes shorter sequences to match the fixed input length by appending neutral values, thereby standardizing input dimensions. This step enhances the model’s ability to analyze and recognize patterns within the data, ultimately improving detection accuracy and performance.

\subsubsection{Ensemble Model}

\begin{figure}[htbp]
\centerline{\includegraphics[width=1\linewidth, height=17cm]{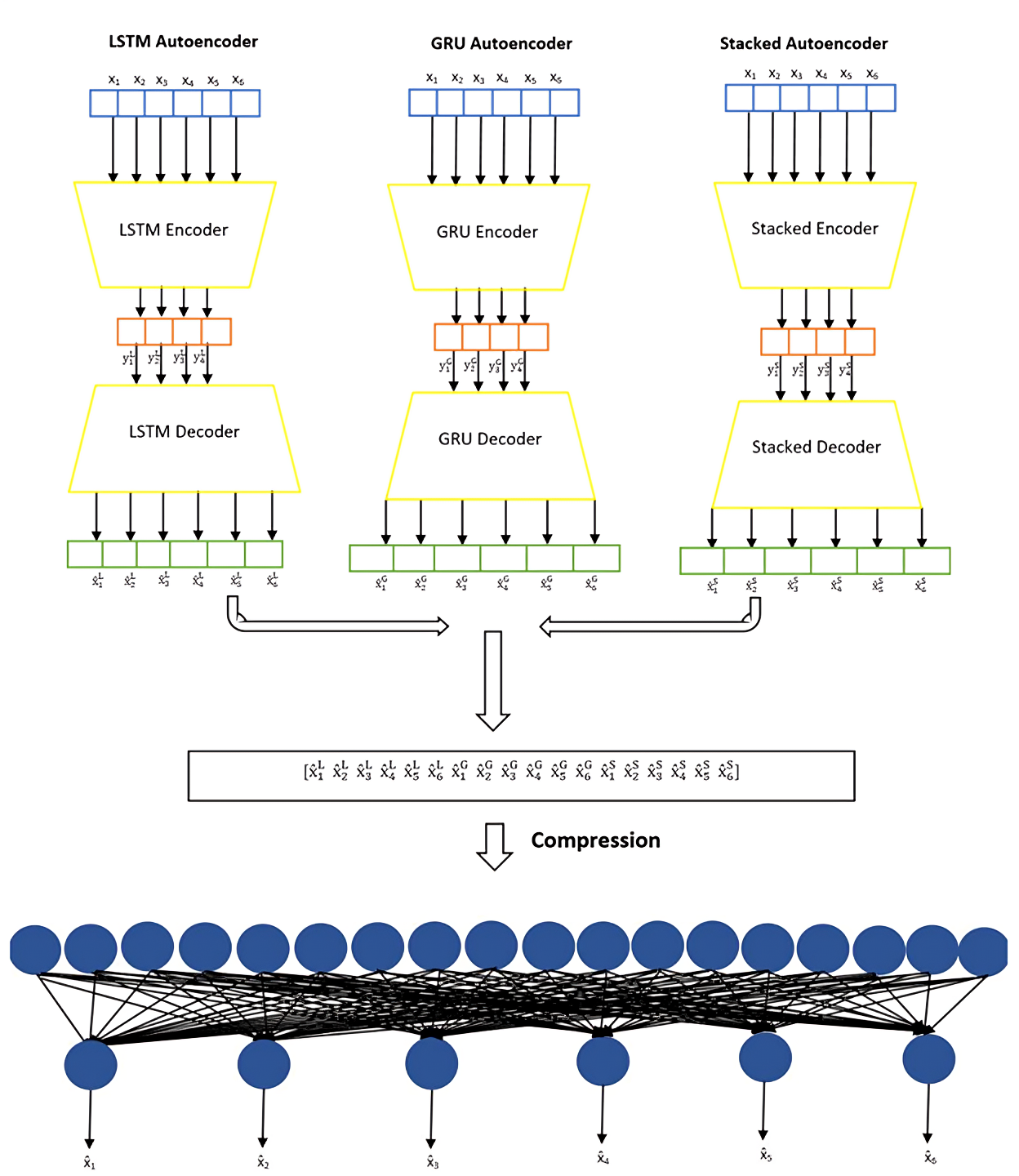}} 
\caption{Structure of the proposed model.}
\label{fig5}
\end{figure}

As illustrated in Figure~\ref{fig5}, the proposed ensemble model consists of three sub-models: an LSTM autoencoder, a GRU autoencoder, and a stacked autoencoder. The initial phase of the proposed approach involves selecting appropriate sub-models for training and detection, ensuring an optimal balance between detection accuracy and computational efficiency. The primary objective is to design multiple lightweight sub-models that, when combined, enhance the overall detection capability while maintaining a low false positive rate.

To achieve this, various neural network architectures were evaluated. Experimental results demonstrated that employing Long Short-Term Memory (LSTM), Gated Recurrent Unit (GRU), and stacked autoencoders in an ensemble configuration enhances data processing efficiency and improves the accurate identification of both known and zero-day web attacks. The outputs of these sub-models are concatenated and further processed through a dense layer for feature reduction, optimizing the final classification process.

Autoencoders are widely utilized for dimensionality reduction and feature extraction. These models consist of an encoder and a decoder, both comprising multiple layers, which collectively transform an input sequence of symbols (words) into a continuous latent representation. The decoder then reconstructs the original input from this representation, preserving critical features while filtering out noise~\cite{vaswani2017attention,rashvand2024enhancing}. This reconstruction-based learning approach enables autoencoders to effectively capture underlying patterns in web requests, further improving the robustness of the detection framework.

The encoded sequences of the original input data, denoted as:

\[
x = [x_1, x_2, x_3, \dots, x_n]
\]

The encoded sequences are obtained through specific encoding functions corresponding to each autoencoder type. The encoding transformations for the LSTM, GRU, and stacked autoencoders are formally defined as follows:

\begin{equation}
    y_L = E_L(x)
\end{equation}

\begin{equation}
    y_G = E_G(x)
\end{equation}

\begin{equation}
    y_S = E_S(x)
\end{equation}

These equations represent the encoding processes, wherein the input sequence \( x \) is mapped to its corresponding latent representation, effectively capturing essential features for the detection process.

For each type of autoencoder, the encoded representations are structured as follows:

\[
y_L = \left[ y_1^L, y_2^L, y_3^L, \dots, y_m^L \right]
\]

for the LSTM autoencoder,

\[
y_G = \left[ y_1^G, y_2^G, y_3^G, \dots, y_m^G \right]
\]

for the GRU autoencoder, and 

\[
y_S = \left[ y_1^S, y_2^S, y_3^S, \dots, y_m^S \right]
\]

for the stacked autoencoder.

The output of the encoder serves as the input to the decoder, which reconstructs the original input sequence from the encoded representation. The reconstruction process is defined as follows:

\begin{equation}
    \hat{x}_L = D_L(y_L)
\end{equation}

\begin{equation}
    \hat{x}_G = D_G(y_G)
\end{equation}

\begin{equation}
    \hat{x}_S = D_S(y_S)
\end{equation}

where \( D_L \), \( D_G \), and \( D_S \) denote the decoding functions for the LSTM, GRU, and stacked autoencoders, respectively.

The primary objective of the decoding process is to validate the quality and representativeness of the extracted features. The outputs from all three autoencoders are subsequently concatenated to form a unified feature representation, denoted as \( \hat{x} \):

\[
\hat{x} = \left[ x_1^L, x_2^L, x_3^L, x_4^L, x_5^L, x_6^L, x_1^G, x_2^G, x_3^G, x_4^G, x_5^G, x_6^G, x_1^S, x_2^S, x_3^S, x_4^S, x_5^S, x_6^S \right]
\]

To maintain consistency with the original input dimensions and optimize computational efficiency, a compression operation is applied to \( \hat{x} \). This step ensures that the number of features in the final output aligns with that of the original input data, enabling effective processing and interpretation of web requests within the proposed model:

\[
\hat{x} = \left[ \hat{x}_1, \hat{x}_2, \hat{x}_3, \dots, \hat{x}_n \right]
\]

This final transformation refines the extracted feature representations, ensuring that the model retains only the most relevant and informative aspects of the input data while discarding redundant or insignificant components. By optimizing the structure of the encoded sequences, the model enhances its capacity for accurate detection and classification of web requests.

\section{Evaluation and Results}
\label{sec:evaluation_results}

This section presents the evaluation of the proposed model and its sub-models based on multiple performance metrics, including accuracy, detection rate, sensitivity, precision, and false positive rate. To assess the effectiveness of the proposed approach, a threshold-based evaluation is conducted using the Mean Absolute Error (MAE), which quantifies the difference between the reconstructed request and the original input (prior to encoding).

In machine learning, MAE is a widely used metric for measuring the absolute difference between predicted values and their actual counterparts. It is computed by averaging the absolute errors across all predictions. MAE was selected as the primary evaluation metric due to its interpretability, robustness, and alignment with the model’s objectives. Specifically, MAE measures the average absolute deviation between the original and reconstructed web requests, providing a clear and intuitive indicator of reconstruction accuracy.

Unlike Mean Squared Error (MSE), which disproportionately amplifies the effect of outliers due to its squared loss formulation, MAE is less sensitive to extreme deviations. This stability makes MAE particularly well-suited for anomaly detection, as it emphasizes individual discrepancies without being overly influenced by rare, extreme variations. By leveraging linear reconstruction errors, MAE effectively differentiates between benign and malicious web requests while ensuring an optimal balance between detection rates and false positives. This makes it an appropriate choice for evaluating the performance of web attack detection models.

The Mean Absolute Error (MAE) is formally defined as:

\begin{equation}
    MAE = \frac{1}{n} \sum_{i=1}^{n} \left| \hat{x}_i - x_i \right|
\end{equation}

\begin{figure}[htbp]
\centerline{\includegraphics[width=1\linewidth, height=9cm]{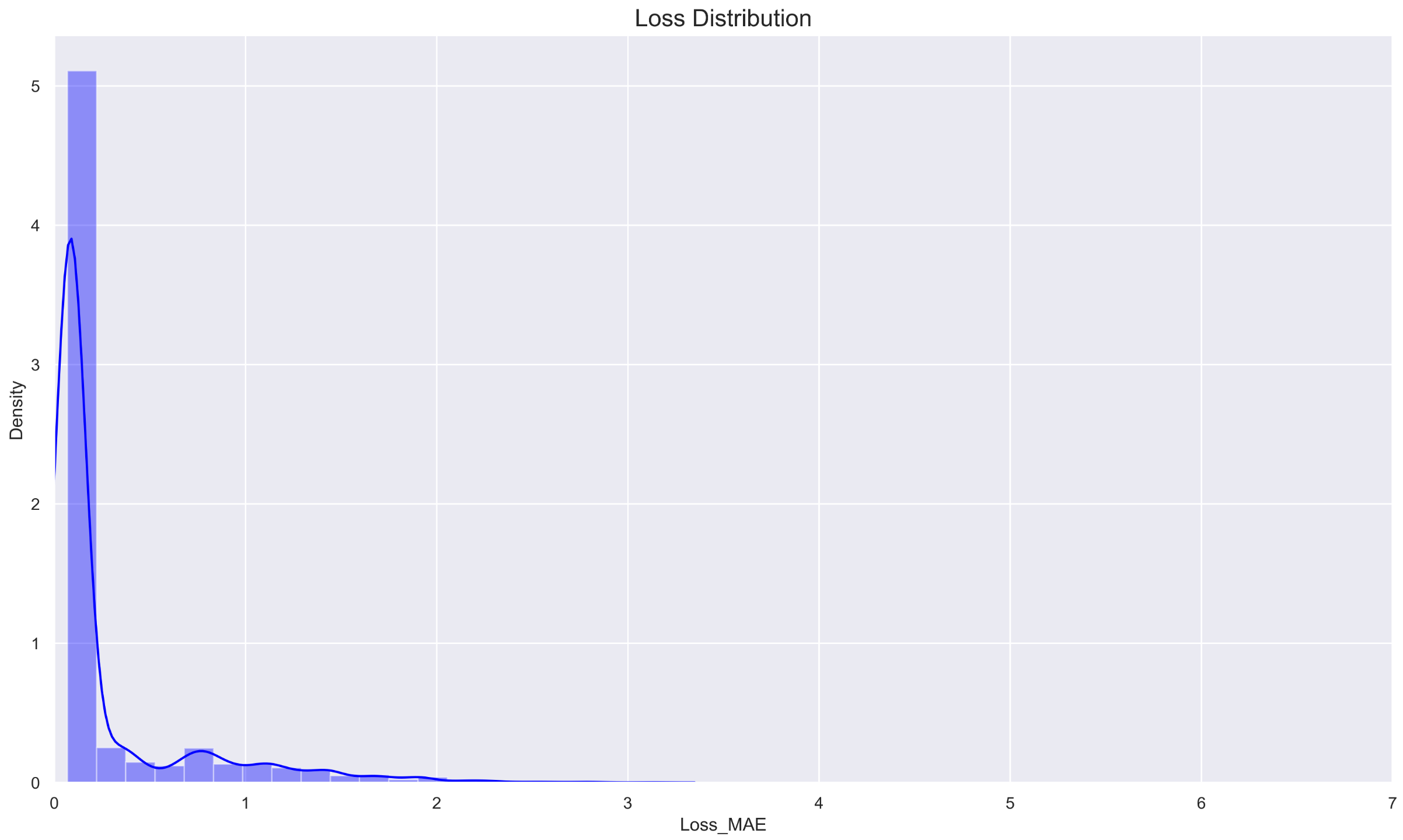}} 
\caption{Density diagram according to MAE of the proposed model.}
\label{fig6} 
\end{figure}

In Formula (7), \( \hat{x}_i \) represents the reconstructed value, while \( x_i \) denotes the actual value. The classification criterion is based on the computed MAE for a given web request: if the MAE falls below a predefined threshold, the request is classified as normal; otherwise, it is identified as malicious. The threshold value is determined using Figure~\ref{fig6}, which visualizes the density distribution of web requests based on the MAE metric. This methodology effectively differentiates benign from malicious web requests, utilizing MAE as a robust anomaly detection measure. Based on the analysis in Figure~\ref{fig6}, an optimal threshold value of approximately 4 is identified.

During training, the neural network is trained exclusively on 80\% of normal web requests, while malicious requests are introduced only during the detection phase. The optimal threshold is determined empirically through iterative experimentation. The analysis suggests that a threshold value of 4.09 provides optimal detection performance, as requests with an MAE of 5 or higher are observed infrequently. However, setting the threshold too high (e.g., at 7) may improve training results but could fail to detect certain malicious requests with reconstruction errors in the range of 5 to 7. This would compromise the model's effectiveness in distinguishing anomalous web activity. A practical example from the model's output further illustrates this decision-making process.

\subsection{Data Collection}

Previous research on detecting malicious and zero-day web requests has primarily relied on two well-established datasets: CSIC~\cite{shaheed2022web} and HTTPPARAMS~\cite{shaheed2022web}. These datasets provide a comprehensive representation of both normal and malicious web requests, making them widely used benchmarks in web security research. Additionally, a project hosted on GitHub~\cite{DuckDuckBugCNNWAF}, which employs Convolutional Neural Networks (CNNs), has utilized the CSIC 2012 dataset. Accordingly, the proposed model leverages this dataset for training and evaluation.

The dataset utilized in this study is significant due to the following characteristics:

\begin{itemize}
    \item It encompasses a diverse range of malicious requests, including SQL Injection (SQLi), Cross-Site Scripting (XSS), and Buffer Overflow attacks.
    \item It contains normal (benign) web requests, ensuring a balanced distribution of data for effective training and evaluation.
\end{itemize}

A key consideration in dataset selection is ensuring that malicious requests accurately reflect real-world attack scenarios. The dataset comprises approximately 16,000 instances labeled as anomalous. However, certain anomalies may arise from factors unrelated to direct cyberattacks, such as unusual user behavior, malformed requests, or suspicious data entry attempts. These cases, while indicative of potential security threats, do not strictly conform to defined attack patterns. To maintain data integrity and ensure the model is trained on well-defined attack and normal request samples, such ambiguous anomalies are removed from the dataset prior to training.

\subsection{The Ensemble Model Structure}

\begin{figure}[htbp]
\centerline{\includegraphics[width=1.1\linewidth, height=8cm]{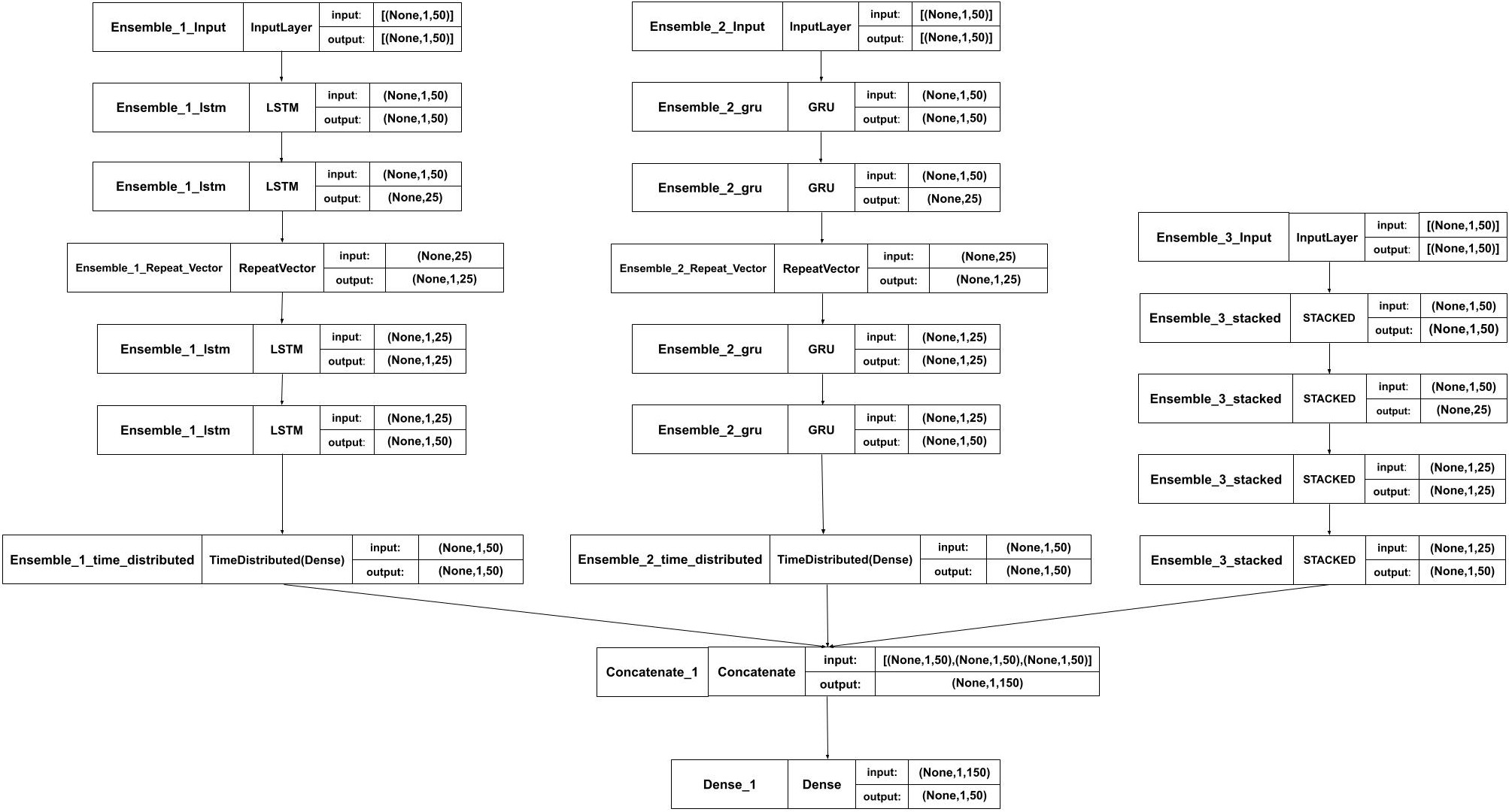}} 
\caption{Ensemble model architecture.}
\label{fig9} 
\end{figure}

According to Figure~\ref{fig9} the LSTM autoencoder and GRU autoencoder each consist of four layers (two encoder layers of 50 and 25 units respectively, and two symmetric decoder layers of 25 and 50 units), using the default tanh activation function. The stacked autoencoder comprises four dense layers (50, 25, 25, and 50 units) with linear activation. The ensemble model concatenates outputs from these autoencoders into a unified latent vector, which is further compressed via a dense layer (50 units). All models are trained using the Mean Absolute Error (MAE) loss function, Nadam optimizer, and evaluated based on accuracy metrics.

\begin{figure}[htbp]
\centerline{\includegraphics[width=1\linewidth, height=7cm]{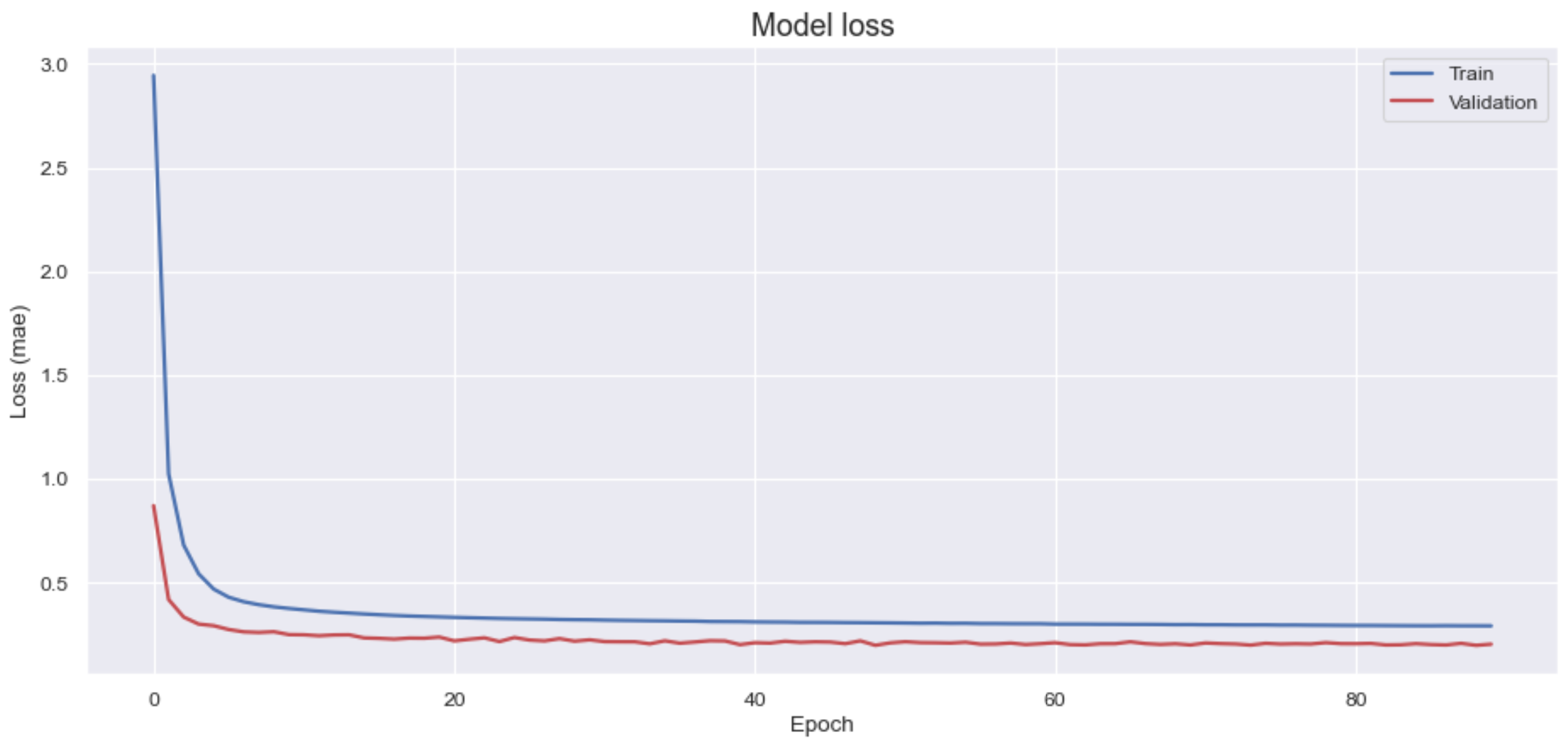}} 
\caption{Training Process of the Proposed Model Based on MAE and Number of Epochs.}
\label{fig7} 
\end{figure}

The evaluation results of the proposed model, as depicted in Figure~\ref{fig7}, illustrate the training process. Over the course of 90 epochs, the Mean Absolute Error (MAE) metric exhibited a significant reduction, decreasing from approximately 3 to around 0.5. This trend indicates that, in the initial training stages, a substantial discrepancy existed between the decoded and original values. However, as the ensemble model progressively learned the patterns of normal web requests, the reconstruction error diminished, reflecting an improvement in model accuracy.

The dataset utilized for training and validation followed an 80-20 split, with 80\% of the data allocated for training and the remaining 20\% reserved for validation. This partitioning ensures that the model is effectively trained while being rigorously evaluated on an unseen subset of data, enabling a more reliable assessment of its generalization capability.

\begin{figure}[htbp]
\centerline{\includegraphics[width=1\linewidth, height=7cm]{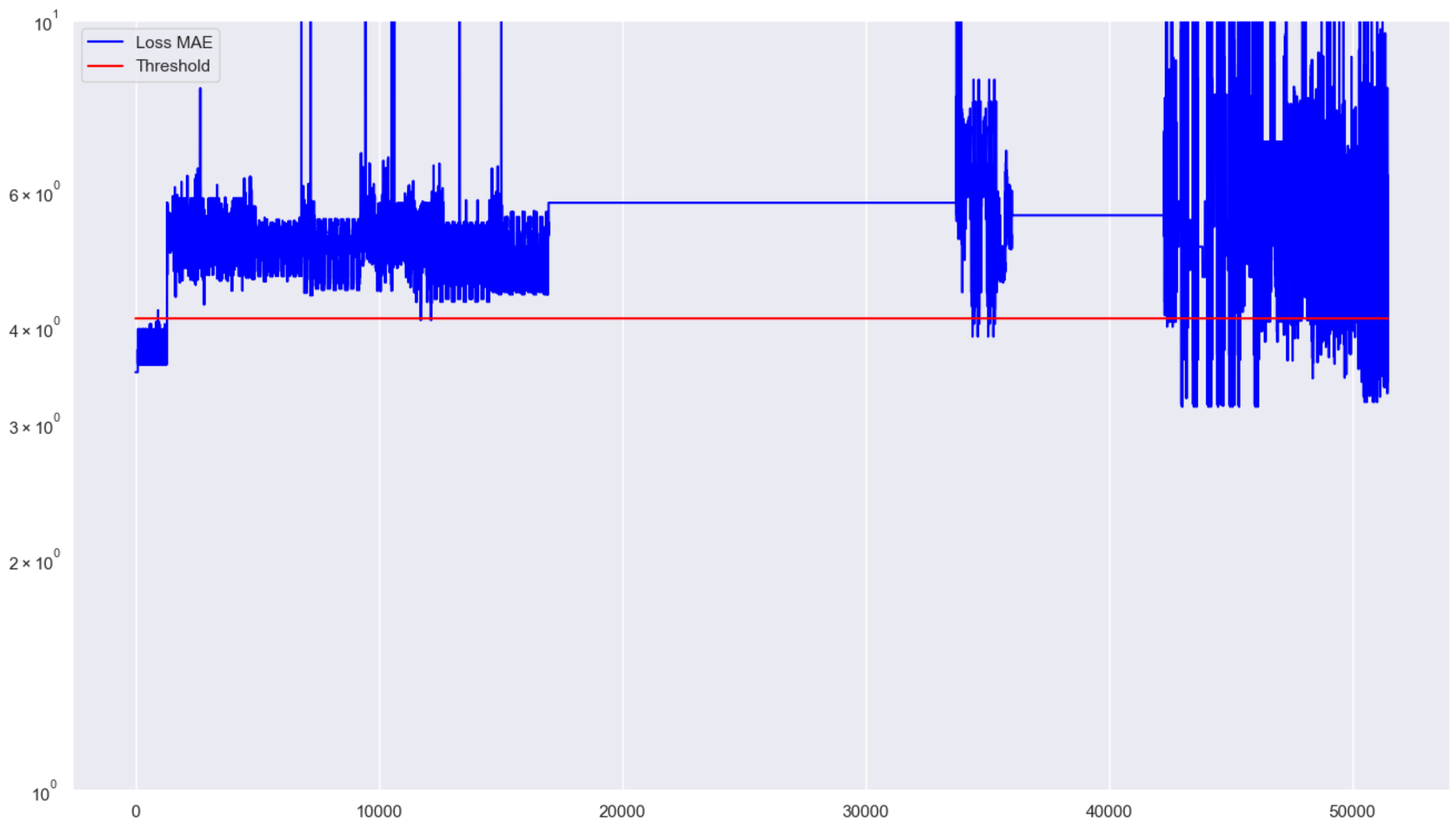}} 
\caption{Comparison of the Calculated MAE Value for Each Web Request with the Threshold Value.}
\label{fig8} 
\end{figure}

Following the detection phase, the Mean Absolute Error (MAE) for each web request is computed and compared against the predefined threshold. As illustrated in Figure~\ref{fig8}, the calculated MAE values for individual web requests are represented in blue, while the threshold value is indicated by a red line. Web requests with an MAE exceeding the threshold (blue points above the red line) are classified as malicious, whereas those with an MAE below the threshold (blue points below the red line) are categorized as normal requests.

The ensemble model, incorporating LSTM, GRU, and stacked autoencoder sub-models, demonstrates superior performance across all evaluation metrics compared to each sub-model individually. The reported results represent the average performance obtained over six independent runs of the model. 

The system utilized for evaluating the proposed model consists of key components, as shown in Table~\ref{tab:system_table}, including LSTM, GRU, and Stacked Autoencoder for neural network-based processing, running on Windows 11 with Python v3.12 as the programming language. The implementation leverages Scikit-learn v1.6.0 for machine learning functionalities and WordPunctTokenizer from the Natural Language Toolkit (NLTK) for splitting a text into a sequence of words. Additionally, the Tokenizer class is employed for converting text data into numerical sequences, ensuring compatibility with neural networks. The model’s performance is evaluated using Mean Absolute Error (MAE), as previously defined, to quantifies the difference between predicted and actual values, providing an effective measure for anomaly detection. The training phase required approximately 20 seconds, while the test phase was completed in 5 seconds. The model itself was implemented using Python version 3.12 with the Keras framework.

\begin{table}[h]
    \centering
    \caption{System components used in the evaluation setup.}
    \label{tab:system_table}
    \renewcommand{\arraystretch}{1.2} 
    \begin{tabular}{|p{4cm}|p{7cm}|} 
        \hline
        \textbf{System} & \textbf{Details} \\ 
        \hline
        Neural Networks & LSTM, GRU, and Stacked Autoencoder \\ 
        \hline
        Operating System & Windows 11 \\ 
        \hline
        Programming Language & Python v3.12 \\ 
        \hline
        Python Library & Scikit-learn v1.6.0 \\ 
        \hline
        Natural Language Toolkit (NLTK) Library & WordPunctTokenizer  \\ 
        \hline
        Feature Extraction and Tokenization Tool & Tokenizer class in python \\ 
        \hline
        Evaluation Metric for measuring prediction accuracy & Mean Absolute Error (MAE) \\ 
        \hline
    \end{tabular}
\end{table}

\subsection{Results}

Table~\ref{tab:metrics_table} defines the key terms used to compute the evaluation metrics. The performance assessment of the proposed model involves the computation of six primary metrics: accuracy, precision, sensitivity, detection rate, false positive rate, and F1 score.

The proposed ensemble model consistently outperforms the individual sub-models across all evaluation metrics. While the LSTM and GRU autoencoders achieve high accuracy, sensitivity, and precision, they exhibit a higher false positive rate, incorrectly classifying several normal requests as malicious. Conversely, the stacked autoencoder reduces the false positive rate effectively but shows comparatively weaker precision and recall. Combining these sub-models into a unified ensemble framework leverages their complementary strengths, thereby significantly improving overall detection performance.

\begin{figure}[htbp]
\centerline{\includegraphics[width=1.1\linewidth, height=9cm]{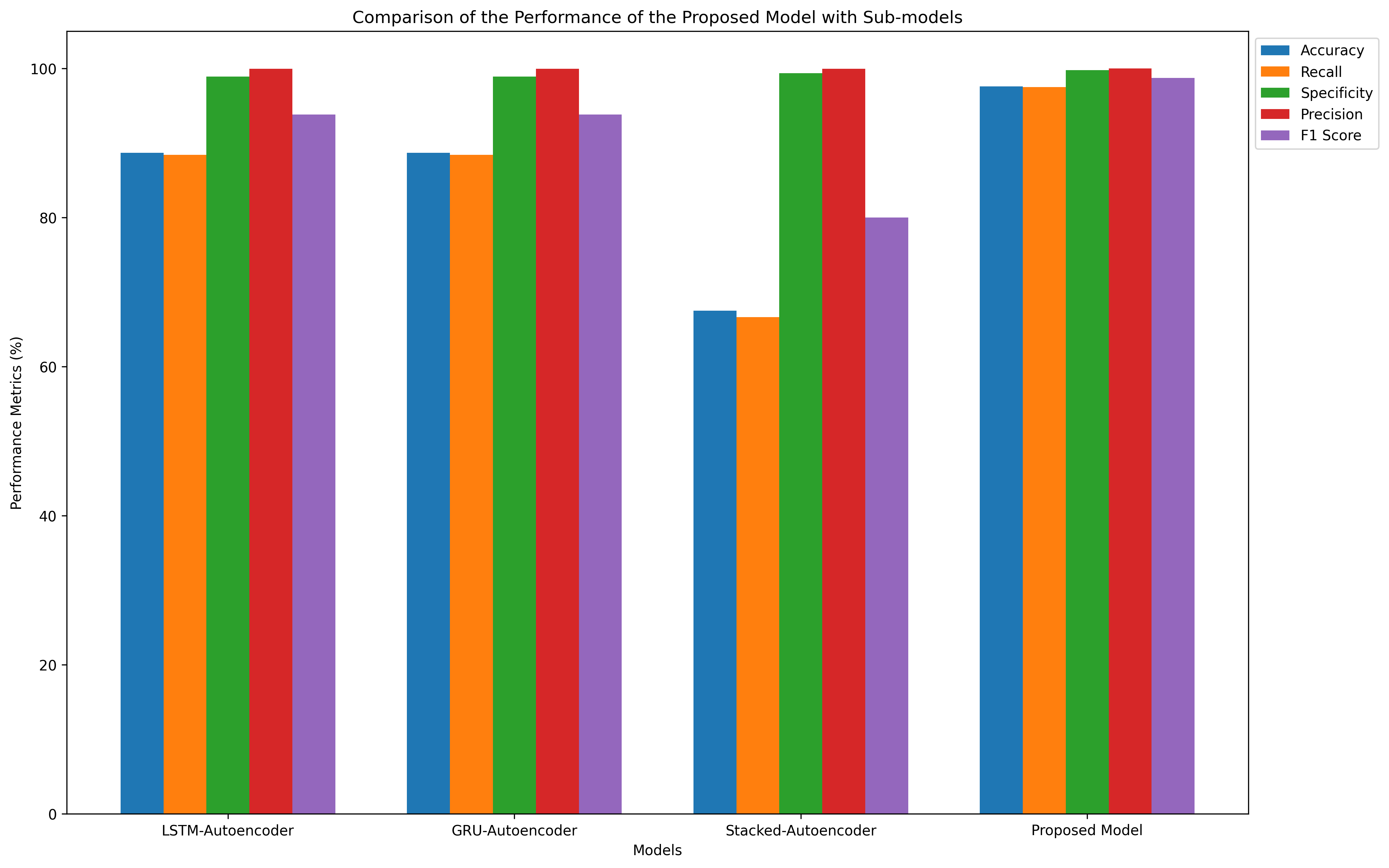}} 
\caption{Comparison of the Performance of the Proposed Model with the Sub-models in the form of a bar chart.}
\label{fig10} 
\end{figure}

A detailed analysis reveals that both LSTM and GRU sub-models misclassified 14 out of 1,299 normal requests as malicious—an undesirable outcome in real-world scenarios. Incorporating the stacked autoencoder into the ensemble mitigates this issue by reducing false positives, albeit at the expense of slightly lower accuracy and recall when used independently. Table~\ref{tab:performance_comparison} clearly illustrates the comparative performance of each individual sub-model against the proposed ensemble approach and Figure~\ref{fig10} illustrates this comparison in the form of a bar chart.

\begin{table}[h]
    \centering
    \caption{Performance metrics used for the proposed model.}
    \label{tab:metrics_table}
    \renewcommand{\arraystretch}{1.3}
    \begin{tabular}{|l|l|c|}
        \hline
        \textbf{Metric} & \textbf{Definition/Calculation} & \textbf{Value} \\
        \hline
        Total (T) & Total number of requests & 51473 \\
        \hline
        Correct Predictions (CP) & Number of correctly predicted requests & 50230 \\
        \hline
        Negatives & Normal requests & 1299 \\
        \hline
        Positives & Malicious requests & 50174 \\
        \hline
        True Negatives (TN) & Normal requests correctly predicted as normal & 1296 \\
        \hline
        False Positives (FP) & Normal requests incorrectly predicted as malicious & 3 \\
        \hline
        False Negatives (FN) & Malicious requests incorrectly predicted as normal & 1240 \\
        \hline
        True Positives (TP) & Malicious requests correctly predicted as malicious & 48934 \\
        \hline
        Accuracy & \( \frac{TN + TP}{TP + TN + FP + FN} \) & 0.9758 \\
        \hline
        Recall (Sensitivity) & \( \frac{TP}{TP + FN} \) & 0.9752 \\
        \hline
        Specificity & \( \frac{TN}{TN + FP} \) & 0.9976 \\
        \hline
        Precision & \( \frac{TP}{TP + FP} \) & 0.9999 \\
        \hline
        False Positive Rate & \( 1 - \text{Specificity} \) & 0.002 \\
        \hline
        F1 Score & \( 2 \times \frac{\text{Precision} \times \text{Recall}}{\text{Precision} + \text{Recall}} \) & 0.9874 \\
        \hline
    \end{tabular}
\end{table}

These results demonstrate that the ensemble strategy effectively balances the distinct advantages and limitations of each sub-model, providing a robust and reliable solution for detecting zero-day web attacks.

\begin{table}[h]
    \centering
    \caption{Comparison of the Performance of the Proposed Model with the Sub-models.}
    \label{tab:performance_comparison}
    \renewcommand{\arraystretch}{1.3} 
    \resizebox{\textwidth}{!}{ 
    \begin{tabular}{|p{3.5cm}|p{2cm}|p{2cm}|p{2cm}|p{2cm}|p{2.5cm}|p{2cm}|}
        \hline
        \textbf{Models} & \textbf{Accuracy} & \textbf{Recall} & \textbf{Specificity} & \textbf{Precision} & \textbf{False Positive Rate} & \textbf{F1 Score} \\ 
        \hline
        LSTM-Autoencoder & 88.68\% & 88.41\% & 98.92\% & 99.96\% & 1\% & 93.83\% \\ 
        \hline
        GRU-Autoencoder & 88.69\% & 88.42\% & 98.92\% & 99.96\% & 1\% & 93.83\% \\ 
        \hline
        Stacked-Autoencoder & 67.48\% & 66.65\% & 99.38\% & 99.97\% & 0.6\% & 79.98\% \\ 
        \hline
        \textbf{Proposed Model} & \textbf{97.58\%} & \textbf{97.52\%} & \textbf{99.76\%} & \textbf{99.99\%} & \textbf{0.2\%} & \textbf{98.74\%} \\ 
        \hline
    \end{tabular}
    }
\end{table}

\begin{figure}[htbp]
\centerline{\includegraphics[width=1.1\linewidth, height=8cm]{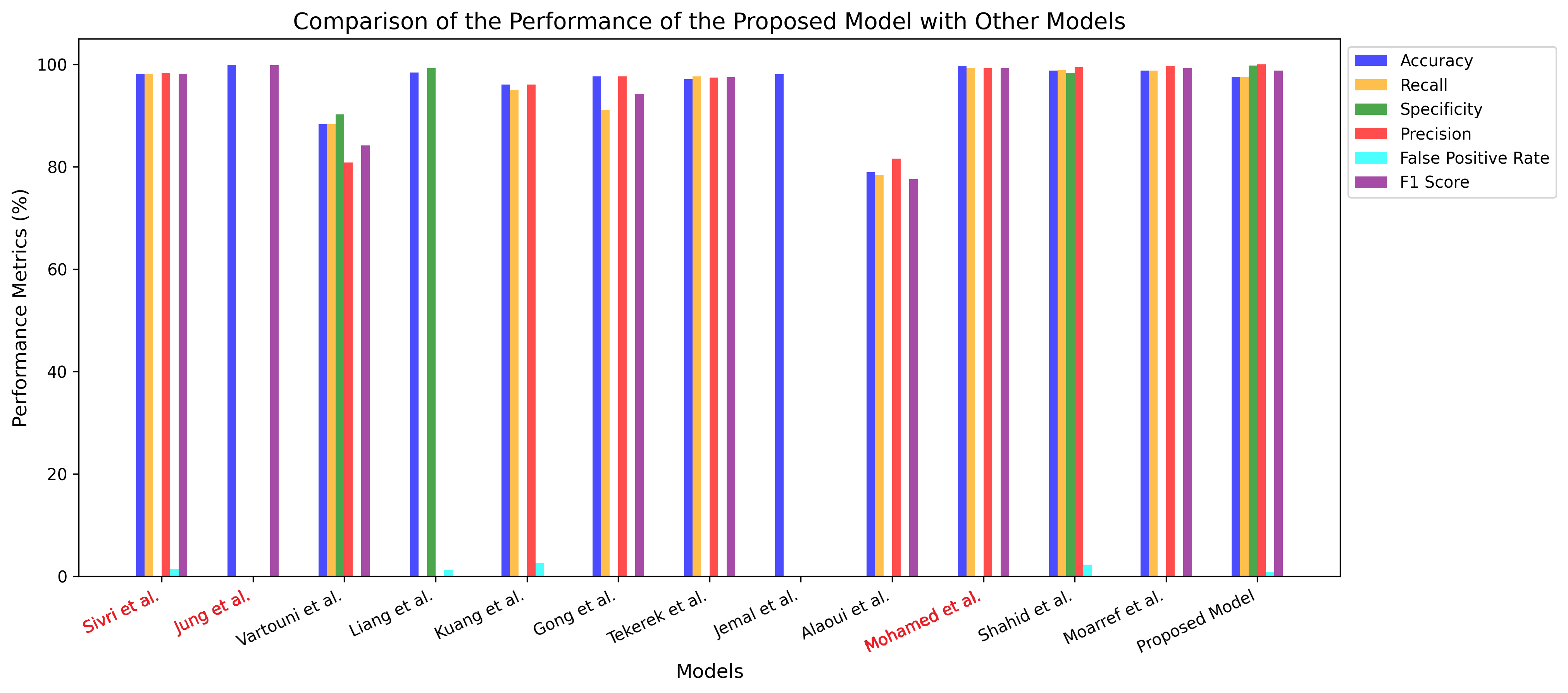}} 
\caption{Comparison of the Performance of the Proposed Model with Previously Designed Models in the form of a bar chart.}
\label{fig11} 
\end{figure}

Table~\ref{tab:comparison_previous_models} presents a comparative analysis of the proposed model's performance against various models from previous research that have utilized the CSIC2010 and CSIC2012 datasets. This comparison provides insights into the effectiveness of the proposed approach relative to existing solutions in the field. One notable limitation in prior studies is the omission of the False Positive Rate (FPR) in their evaluation results. This metric is crucial, as it quantifies the number of normal requests misclassified as malicious, directly impacting the practical applicability of detection models. Figure~\ref{fig11} illustrates this comparison in the form of a bar chart. In the figure, models that utilized the CSIC2012 dataset are highlighted in red.

The primary comparison focuses on studies that have employed the CSIC2012 dataset~\cite{sivri2022web,jung2021pf,mohamed2023multi}, as they provide the most directly comparable benchmark. However, to offer a broader perspective, we also include studies based on the CSIC2010 dataset~\cite{vartouni2018anomaly,liang2017anomaly,kuang2019deepwaf,gong2021model,tekerek2021novel,jemal2022swaf,alaoui2022web,shahid2022enhanced,moarref2023mc}. It is important to note that differences in dataset characteristics may influence the comparability of results.

The CSIC2010 and CSIC2012 datasets are widely recognized benchmarks for evaluating web application security models, particularly for detecting SQL Injection (SQLi) and other web-based attacks. The CSIC2010 dataset, developed earlier, contains a diverse set of normal and anomalous HTTP requests. While it provides a solid foundation for studying web attack detection, it lacks the complexity and evolving attack patterns characteristic of modern cybersecurity threats. 

To address these limitations, the CSIC2012 dataset was designed with more sophisticated and realistic attack scenarios, along with a broader range of normal traffic. This makes CSIC2012 a more representative dataset for contemporary web security challenges. Additionally, CSIC2012 includes refined labeling and a larger volume of data, enhancing its suitability for training and evaluating advanced machine learning models.

These distinctions underscore the importance of selecting CSIC2012 for research targeting modern web application threats, as it serves as a more rigorous and up-to-date evaluation benchmark compared to its predecessor.

\begin{table}[h]
    \centering
    \caption{Comparison of the Performance of the Proposed Model with Previously Designed Models.}
    \label{tab:comparison_previous_models}
    \renewcommand{\arraystretch}{1.3} 
    \resizebox{\textwidth}{!}{ 
    \begin{tabular}{|p{5cm}|p{2cm}|p{2cm}|p{2cm}|p{2cm}|p{2.5cm}|p{2cm}|}
        \hline
        \textbf{Models} & \textbf{Accuracy} & \textbf{Recall} & \textbf{Specificity} & \textbf{Precision} & \textbf{False Positive Rate} & \textbf{F1 Score} \\ 
        \hline
        Sivri et al. \cite{sivri2022web} & 98.15\% & 98.15\% & - & 98.20\% & 0.8\% & 98.16\% \\ 
        \hline
        Jung et al. \cite{jung2021pf} & 99.88\% & - & - & - & - & 99.80\% \\ 
        \hline
        Vartouni et al. \cite{vartouni2018anomaly} & 88.32\% & 88.34\% & 90.20\% & 80.79\% & - & 84.12\% \\ 
        \hline
        Liang et al. \cite{liang2017anomaly} & 98.42\% & - & 99.21\% & - & 0.7\% & - \\ 
        \hline
        Kuang et al. \cite{kuang2019deepwaf} & 96\% & 95\% & - & 96\% & 2\% & - \\ 
        \hline
        Gong et al. \cite{gong2021model} & 97.64\% & 91.11\% & - & 97.62\% & - & 94.25\% \\ 
        \hline
        Tekerek et al. \cite{tekerek2021novel} & 97.07\% & 97.59\% & - & 97.43\% & - & 97.51\% \\ 
        \hline
        Jemal et al. \cite{jemal2022swaf} & 98.1\% & - & - & - & - & - \\ 
        \hline
        Alaoui et al. \cite{alaoui2022web} & 78.95\% & 78.41\% & - & 81.54\% & - & 77.57\% \\ 
        \hline
        Mohamed et al. \cite{mohamed2023multi} & 99.66\% & 99.28\% & - & 99.18\% & - & 99.22\% \\ 
        \hline
        Shahid et al. \cite{shahid2022enhanced} & 98.73\% & 98.87\% & 98.33\% & 99.41\% & 1.67\% & 99.13\% \\ 
        \hline
        Moarref et al. \cite{moarref2023mc} & 99.36\% & 98.80\% & - & 99.65\% & - & 99.22\% \\ 
        \hline
        \textbf{Proposed Model} & \textbf{97.58\%} & \textbf{97.52\%} & \textbf{99.76\%} & \textbf{99.99\%} & \textbf{0.2\%} & \textbf{98.74\%} \\ 
        \hline
    \end{tabular}
    }
\end{table}

\section{Discussion} 
\label{sec:discussion}
According to the obtained results, the proposed model demonstrates strong performance across all evaluation metrics, particularly in terms of the False Positive Rate (FPR). The model achieves the lowest FPR compared to related work referenced in this study, highlighting its ability to accurately differentiate between normal and malicious web requests. This is a critical aspect of web security, as a high FPR would indicate an increased likelihood of blocking legitimate users, thereby compromising usability.

\subsection{Effectiveness of the Ensemble Model}

The proposed ensemble model comprises three fundamental sub-models: LSTM, GRU, and Stacked Autoencoder. Each sub-model plays a distinct role in detecting malicious web requests, contributing unique advantages:

\begin{itemize}
    \item \textbf{LSTM Autoencoder:} Captures long-term dependencies in web request sequences, enhancing the ability to recognize complex request structures.
    \item \textbf{GRU Autoencoder:} Provides computational efficiency while preserving strong sequential pattern recognition capabilities.
    \item \textbf{Stacked Autoencoder:} Focuses on dimensionality reduction and latent feature extraction, improving anomaly detection in subtle attack patterns.
\end{itemize}

By integrating these sub-models, the ensemble model achieves a well-balanced performance across multiple evaluation metrics, effectively mitigating the weaknesses of individual sub-models by an innovative concatenation and feature-compression mechanism. This structured ensemble method distinctly differs from traditional ensemble strategies and clearly enhances computational efficiency and detection accuracy. Additionally, our advanced tokenization method provides a unique and consistent input representation that significantly improves anomaly detection compared to conventional tokenization strategies. Our explicit evaluation of the False Positive Rate further distinguishes our work by providing practical insights often neglected in related literature. The ensemble approach significantly enhances accuracy, recall, and precision while substantially reducing false positives compared to each sub-model independently.

\subsection{Practical Considerations and Limitations}

Although the proposed model demonstrates robust performance on benchmark datasets, its real-world deployment in web security applications requires additional considerations. One of the primary challenges is the variability in web request patterns across different applications and domains. While the tokenization and feature extraction methods effectively standardize the CSIC2012 dataset, these techniques should be adapted and applied to other datasets, such as FWAF and HTTPParams~\cite{jagat2024detecting}, to enhance generalization.

Additionally, while the model is designed to detect zero-day attacks, it does not explicitly classify attack types (e.g., SQL Injection, Cross-Site Scripting (XSS), Buffer Overflow). Future research should focus on integrating an attack-type classification mechanism alongside anomaly detection to improve forensic analysis and threat response capabilities.

\section{Conclusions}
\label{sec:conclusions}

In this study, each web request was initially segmented into individual words and then tokenized using a predefined vocabulary. This preprocessing step aimed to standardize and simplify web requests while establishing a structured pattern for normal web traffic. In the final stage of preprocessing, each tokenized word was mapped to a unique numerical representation, facilitating its input into the neural network. The proposed model employs an ensemble approach comprising three relatively simple sub-models: LSTM, GRU, and Stacked Autoencoders. The ensemble operates by independently processing the input data through each sub-model and then outputs are explicitly concatenated into a combined latent feature set, ensuring the ensemble benefits from the diverse representation capabilities of each sub-model.
After concatenation, a dedicated Dense layer compresses the resulting features into a unified, optimized representation, significantly reducing the dimensionality from a larger combined vector to a manageable size. A novel structured tokenization method significantly enhancing detection performance, and explicit evaluation of critical metrics including False Positive Rate.

During the training phase, only normal web requests were provided as input to the ensemble model, enabling it to learn the underlying patterns of legitimate requests. Upon completion of training, the model effectively captured and recognized these patterns. In the detection phase, both normal and malicious web requests were introduced for evaluation. The Mean Absolute Error (MAE) was employed as the primary metric to quantify the difference between the reconstructed and original values of each request. The threshold for classification was determined based on the MAE values computed during the training phase. In the detection phase, if the MAE of a web request was below the threshold, it was classified as normal; otherwise, it was identified as malicious.

During evaluation, the ensemble model's performance was compared against each of its sub-models individually. The results demonstrated that the ensemble approach achieved superior performance, particularly in terms of an increased detection rate and a reduced false positive rate. Additionally, the proposed model was benchmarked against prior research, where it consistently outperformed existing approaches, further validating its effectiveness in detecting web-based threats.

\section{Future Work}
\label{sec:future work}
One potential direction for future research involves enhancing the tokenization and feature extraction process~\cite{abshari2024llm} across diverse web attack datasets. This improvement can be achieved through the application of Generative AI, leveraging Large Language Models (LLMs)~\cite{zibaeirad2024comprehensive,abshari2025survey}. Specifically, prompt engineering~\cite{babaey2025gensqli, babaey2025genxss} can be employed to construct a structured prompt that systematically guides the LLM in preprocessing each dataset sample.

A few-shot learning approach can be utilized to accomplish this by providing representative examples from each dataset, detailing the exact tokenization process required for each sample. This method ensures that the LLM internalizes dataset-specific preprocessing rules while maintaining uniformity across different datasets.

Another approach involves automating the preprocessing pipeline by generating customized script code~\cite{white2023prompt} for each dataset. This method leverages LLMs to autonomously generate preprocessing scripts based on structured prompts that define tokenization rules and feature extraction strategies. While manual scripting was employed in this study for data preprocessing, future research will focus on automating this process using LLMs, thereby minimizing human intervention and improving scalability across diverse datasets.

Beyond preprocessing improvements, future efforts will explore the implementation of advanced neural architectures and anomaly detection approaches, such as Bidirectional LSTM, GRU, and Convolutional Neural Networks (CNNs)~\cite{graves2013hybrid}, to develop a more robust ensemble model for web attack detection. Additionally, feature selection techniques will be applied to retain high-information-value features while eliminating less significant ones, effectively reducing input dimensionality and enhancing computational efficiency. Regarding advanced anomaly detection approaches, For instance,~\cite{zibaeirad2025reasoning} introduced VulnSage, a framework leveraging structured reasoning strategies such as Chain-of-Thought and Think and Verify to improve zero-shot vulnerability detection in software systems. Inspired by this work, future studies could explore the use of similar structured reasoning strategies in our web attack detection framework, potentially improving the detection of complex, multi-component web vulnerabilities by incorporating explicit logical analysis and self-verification mechanisms.

Furthermore, the integration of a voting-based ensemble learning approach will be investigated to improve detection accuracy by combining multiple classifiers. This strategy aims to leverage the strengths of individual models, resulting in a more generalized and resilient web attack detection framework.

\begin{adjustwidth}{-\extralength}{0cm}

\reftitle{References}


\bibliography{references}

\end{adjustwidth}
\end{document}